\documentclass{appolb}
\usepackage{epsfig}

\newcommand{\s}{\\ \vspace*{-3mm} }

\begin{document}

\title{Elements of Physics with a Photon Collider%
}

\author{M.M.~M\"uhlleitner
\address{Laboratoire d'Annecy-Le-Vieux de Physique Th\'eorique, LAPTH,
Annecy-Le-Vieux, France}
\and
P.M.~Zerwas
\address{Deutsches Elektronen-Synchrotron, DESY, Hamburg, Germany}
}
\maketitle
\begin{abstract}
After a brief description of the basic principle of a photon collider, 
we summarize the physics potential of such a facility at high energies.
Unique opportunities are provided in supersymmetric theories for the
discovery of heavy scalar and pseudoscalar Higgs bosons as well as 
selectrons and $e$-sneutrinos.
\end{abstract}
\PACS{12.15.-y, 12.38.-t, 12.60.-i}

\vspace*{-12.5cm}
\mbox{ }\hfill DESY 05-241\\
\mbox{ }\hfill LAPTH-CONF-1127/05\\
\mbox{ }\hfill hep--ph/0511339\\

\vspace*{11cm}
\section{Introduction}
Linear colliders, LC, can be operated in several modes among which the $e \gamma$ 
and $\gamma \gamma$ modes provide a wide field of physics opportunities, 
including unique experimental solutions to fundamental problems \cite{r1,r2}.
By means of Compton back-scattering of laser light, almost the entire
energy of electrons/positrons at a linear collider can be transferred to photons 
\cite{r3} so
that $e \gamma$ and $\gamma \gamma$ processes can be studied for energies
close to the TeV scale. The luminosities are
expected to be about one third of the $e^+e^-$ luminosity in the high
energy regime \cite{r1,r2}. Since the cross sections for $e \gamma$ and
$\gamma \gamma$ processes are in general significantly larger than the
cross sections for $e^+e^-$ annihilation processes, the event rates will
be of similar size in the three LC modes. Various options of choosing the
photon polarizations, circular and linear, allow unique experimental
analyses of particle properties and interactions. \\ 

A rich spectrum of interesting physics problems can be studied experimentally 
at a photon collider operating at energies up to a TeV: \s

{{-- {\underline{Higgs physics:}}}} The formation of Higgs bosons 
in $\gamma \gamma$ collisions can
be used to measure precisely the Higgs-$\gamma\gamma$ coupling \cite{r4,r5}. 
Since photons do not couple directly to neutral Higgs bosons, the coupling is 
mediated by virtual charged particle loops, being sensitive to scales 
potentially far beyond the Higgs mass. In the supersymmetry sector, 
$\gamma \gamma$ formation allows us to generate heavy scalar and pseudoscalar 
Higgs bosons \cite{r6,r7} in a wedge centered around medium $\tan\beta$ 
values, in which no other collider, neither the LHC nor the LC $e^+e^-$ mode, 
gives access to the spectrum of heavy Higgs bosons. Near
degeneracy of the scalar and pseudoscalar Higgs bosons will give rise to
large asymmetry effects in $\gamma$ polarization experiments \cite{r8,r9}
if CP is broken in the Higgs sector. \s

{{-- {\underline{Supersymmetry:}}}} A photon collider provides unique
opportunities also in the genuine supersymmetric particle sector. In $e
\gamma$ collisions the production of selectrons in association with
light neutralinos can give access to selectron masses in excess to pair
production in the LC $e^\pm e^-$ modes \cite{r10}. Similarly the production of
$e$-sneutrinos in association with charginos may be used to study the
properties of sneutrinos \cite{r11}.\s

{{-- {\underline{Static electromagnetic parameters:}}}} The large
production cross sections for charged particles in $\gamma \gamma$
collisions can be exploited to determine their static electromagnetic
parameters with high precision. The measurement of the electromagnetic
multi-pole moments can be performed in a pure electromagnetic environment
without interference with weak effects. Examples are the magnetic dipole
moment of the top quark \cite{rr11A}, and the magnetic dipole moment and electric
quadrupole moment of the charged $W^\pm$ bosons \cite{r12,r13}. \s

{{-- {\underline{QCD:}}}} Among the QCD problems which can be addressed
at a high energy photon collider, two problems are of particular interest.
The total $\gamma \gamma$ fusion cross section to hadrons is built up by a
mixture of non-perturbative and perturbative interactions including the
difficult transition zone between the two regimes. 
Of tantamount importance is the analysis of the
quark and gluon content of the photon {\cite{r14}}-{\cite{r15B}}. These parton
distributions had been predicted to be very different from nucleons, with
characteristic properties governed by asymptotic freedom. \s

{{-- {\underline{Varia:}}}} Experiments at a photon collider open many
other windows for interesting search experiments, extending from heavy
Majorana neutrinos, see e.g. Ref.~\cite{r15C}, to excited electrons. The
key is the character of the photon as an almost pure energy quantum which
is very effective in exciting new degrees of freedom. As a result, new 
particles can be
generated with masses very close to the total energy in the system,
most transparent for the electron excitation tower $\{e^\ast\}$ 
of compositeness models in $e \gamma$ collisions. \s

Some of these elements will be reviewed in the third to fifth section of this
report after the basic principle of photon colliders is briefly described
in the next section. 
%
\section{The basic principle}
In $e^+e^-/e^-e^-$ linear colliders nearly the entire electron/positron
energy can be transferred to photons by Compton back-scattering of laser
light. This method has been proposed 
in Refs.~\cite{r3}. The scheme is based on two elements. Kinematically, by 
energy-momentum conservation, a low-energy laser photon must carry away 
almost the entire energy when scattered backward in a collision with a 
high-energy electron/positron. Dynamically, the cross section is maximal,
due to the u-channel exchange of electrons/positrons,
\begin{equation}
A_u \sim {1}/(u-m^2_e) \sim {1}/(1+\beta_e\cos\theta)
\end{equation}
for back-scattering of the photons.
%
\subsection{Photon Energy}
The spectrum is described in detail by the shape function
\begin{equation}
F(y) = \frac{1}{1-y} + (1-y) - 4r(1-r) + 2 \lambda_e P_c xr(1-2r)(2-y)
\end{equation}
where
\begin{equation}
y = E_\gamma/E_e
\end{equation}
denotes the fraction of energy transferred from the electron/positron to
the photon and $r$ abbreviates the ratio $r = y/(1-y)x$. 
The helicities $\lambda_e$ and $P_c$ refer to the incoming electrons/positrons 
and laser photons, respectively.
The parameter
\begin{equation}
x = 4 E_e \omega_\gamma / m^2_e
\end{equation}
measures the invariant energy [squared] in units of the electron mass, with
$\omega_\gamma$ denoting the laser $\gamma$ energy.
$x$ determines the upper limit of the final photon energy,
\begin{equation}
E^{max}_\gamma = \frac{1}{1+x^{-1}} E_e
\end{equation}
The larger the value for $x$ is chosen the more energy can be transferred
to the photon. However, to suppress electron-positron pair production in
collisions of the high-energy photons, already generated in the Compton process,
with the numerous
left-over laser photons, an upper bound on $x \leq 2/(\sqrt{2}-1) \simeq 4.8$
must be imposed. \\

For typical values $E_e =$ 250 GeV and $\omega_\gamma =$ 1.17 eV more than
80\% of the electron/positron energy can be transferred to the photon,
\begin{equation}
E^{max}_\gamma = 0.82 E_e
\end{equation}
generating a photon-photon invariant energy of more than 800 GeV at a
1~TeV $e^\pm e^-$ collider. \\ 

\begin{figure}[h]
\begin{center}
\epsfig{figure=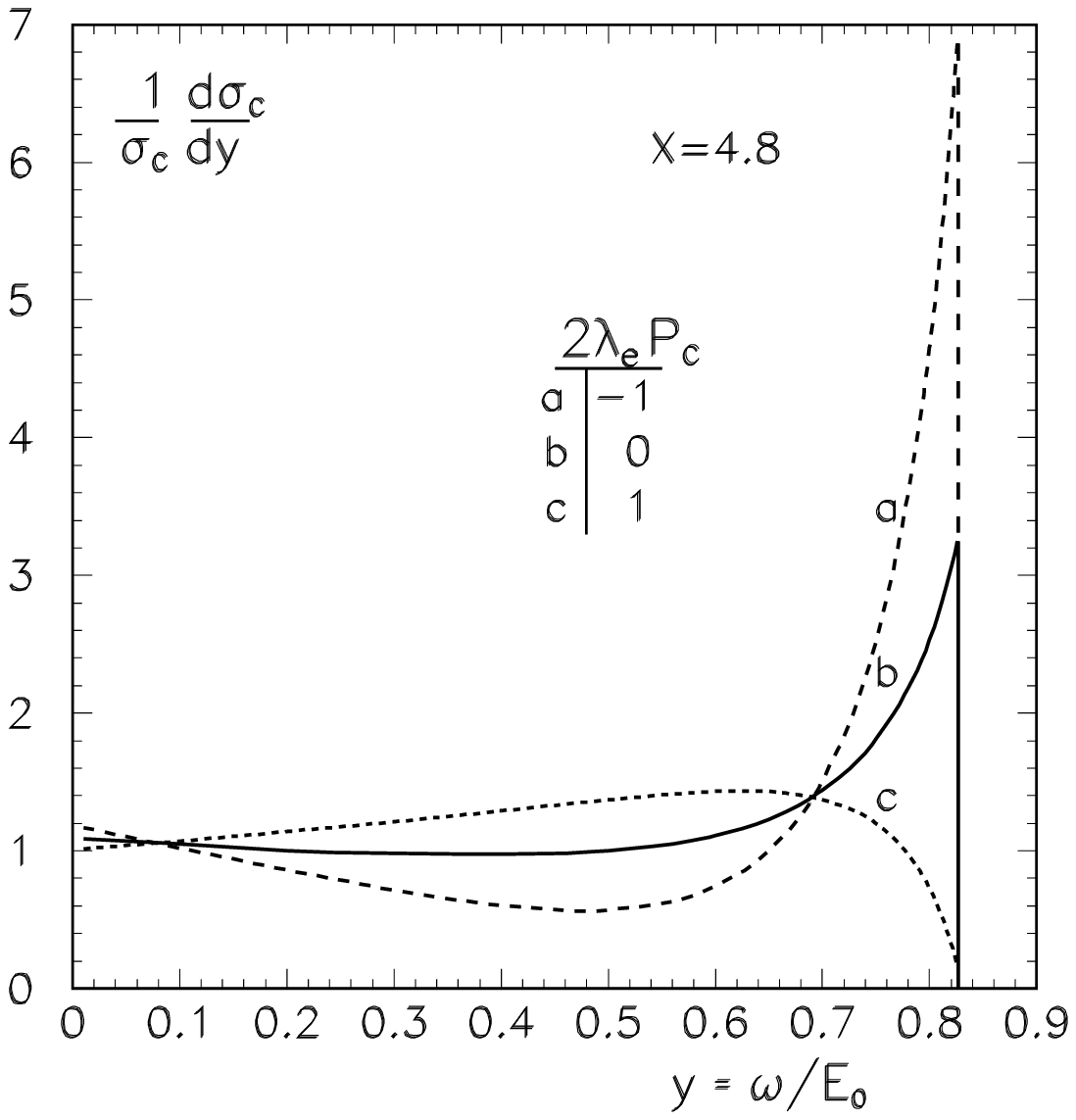,bbllx=63,bblly=45,bburx=381,bbury=375,width=5.1cm,clip=}
\epsfig{figure=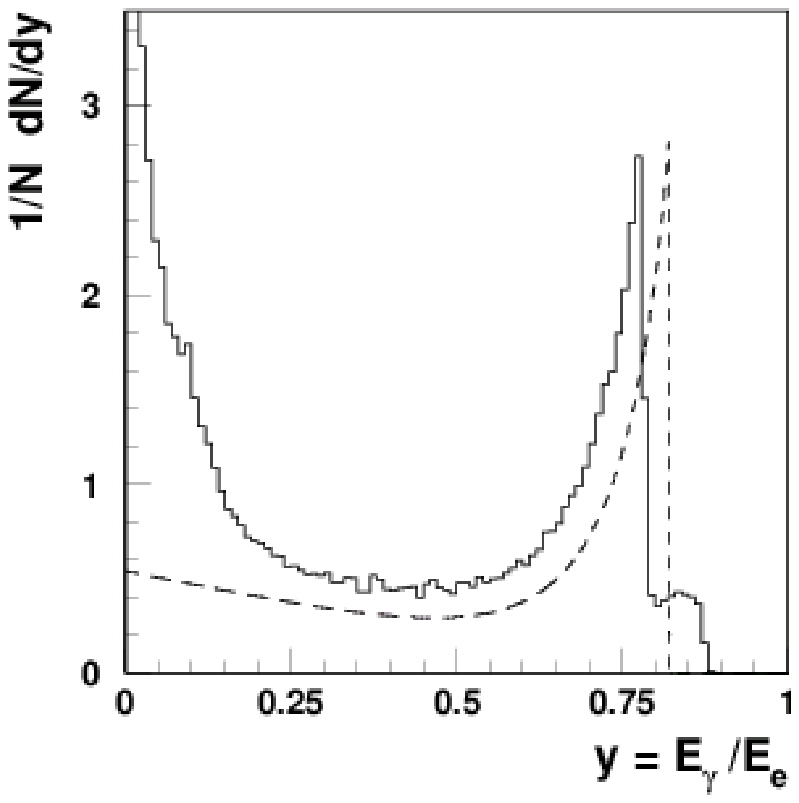,bbllx=3,bblly=6,bburx=238,bbury=245,width=5.4cm,clip=}
\caption{Photon spectra. Left: The effect of electron/positron and laser 
photon polarization on the high-energy photon spectrum. Right: Experimental
simulation of the photon spectrum including non-linear effects; Ref.~\cite{rr17A}.} 
\label{gamspectrum}
\end{center}
\end{figure}

By choosing opposite helicities for the initial electron/positron beams and
the laser photons, i.e. $2\lambda_e P_c = -1$, the spectrum of the final-state 
photons can be sharpened dramatically, cf. Fig.\ref{gamspectrum} (left). The 
$\gamma$ conversion of
electron beams for which a polarization degree of 95\% may eventually be
achieved, is therefore preferable over positron beams where an upper
limit of about 60\% is expected.   

Non-linear effects reduce the upper energy slightly, but give rise to 
a large number of photons at the lower end of the spectrum. Higher harmonics, 
on the other hand, lead to a shallow increase of the upper energy limit, 
cf. Fig.\ref{gamspectrum} (right). The main characteristics of the high-energy 
photon spectrum, however, remain largely unaltered. 
%
\subsection{Polarization}
\begin{figure}[h]
\begin{center}
\epsfig{figure=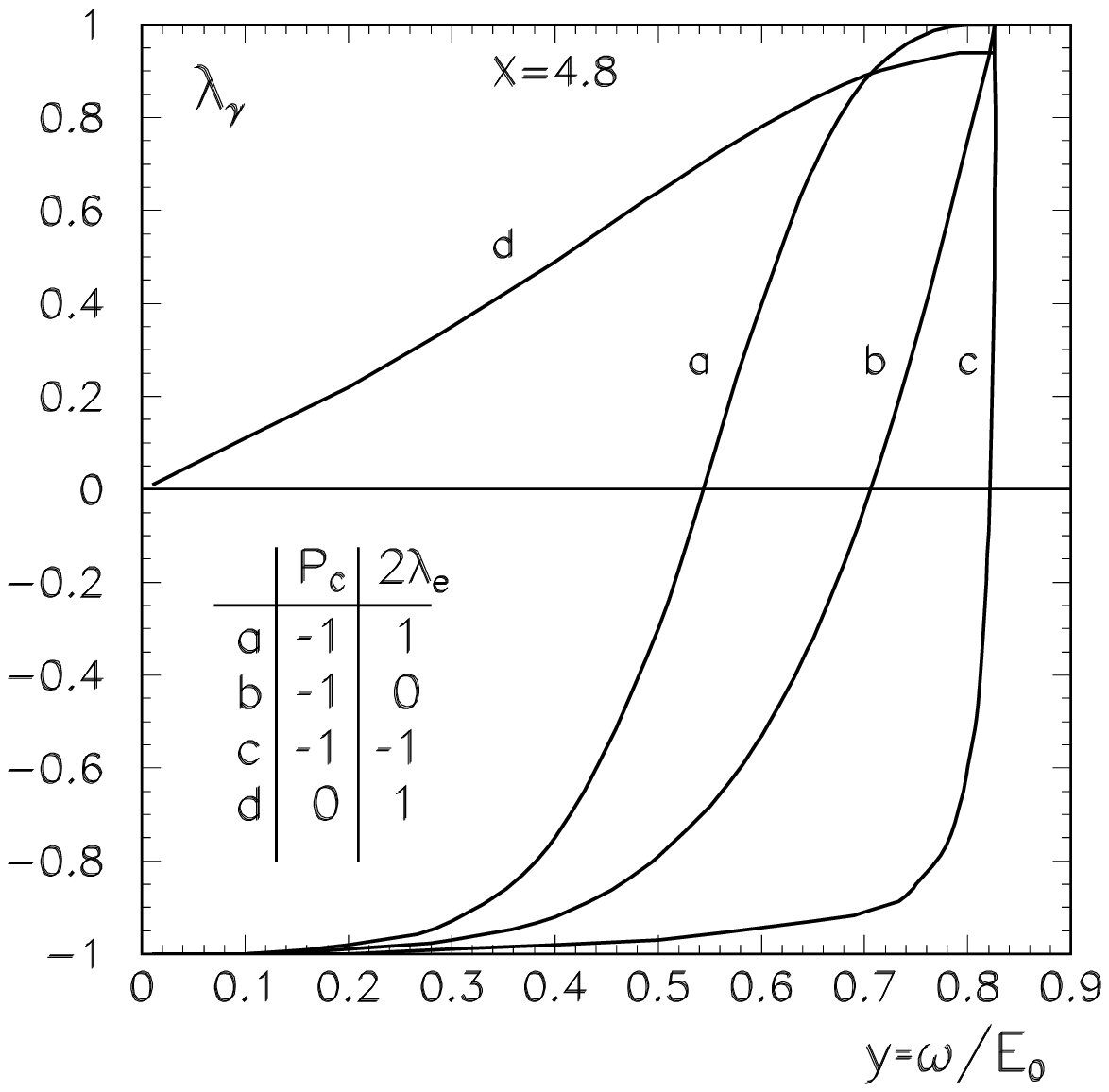,bbllx=41,bblly=43,bburx=381,bbury=371,width=5.4cm,clip=}
\epsfig{figure=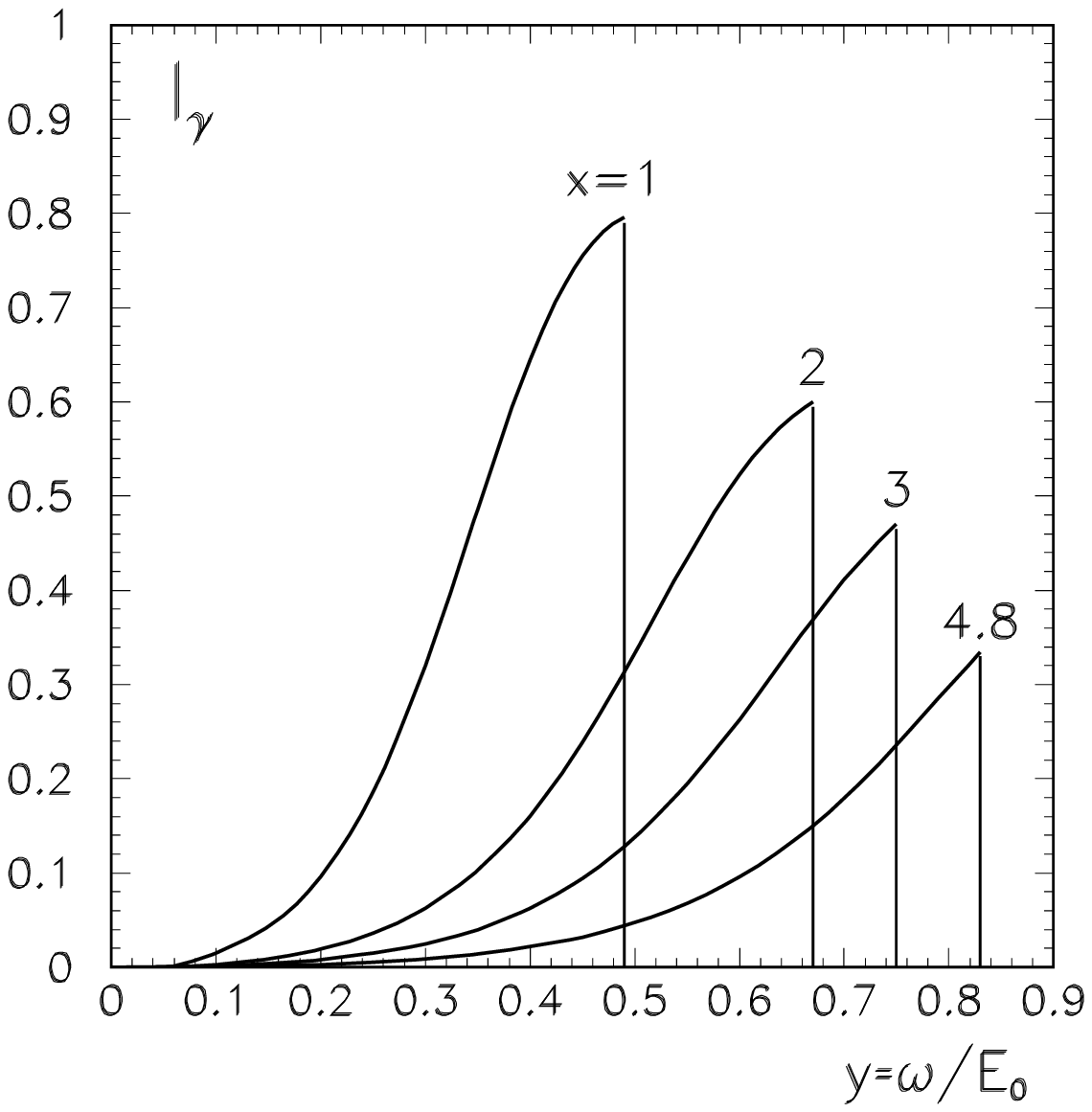,bbllx=41,bblly=43,bburx=381,bbury=371,width=5.4cm,clip=}
\caption{The transfer of circular (left) and linear (right) laser $\gamma$ polarization 
to the high-energy photons.}
\label{circlinspectrum}
\end{center}
\end{figure}

-- {{Circular polarization}} is transferred from the initial laser
photons completely to the high energy photons for maximum energy at the peak 
of the spectrum, cf. Fig.\ref{circlinspectrum} (left):
\begin{equation}
\lambda_\gamma \to - P_c \;\; \mathrm{[near\,maximum]}
\end{equation}

-- {{Linear polarization}} responds less favorably, cf. 
Fig.\ref{circlinspectrum} (right).
The degree of polarization transferred from the laser to the high energy 
photons is reduced with rising photon energy, i.e. rising $x$:
\begin{equation}
l_\gamma \to + P_l/[x^2/2(x+1)+1] \;\; \mathrm{[near\,maximum]}
\end{equation}

The variety of polarization states of the high energy photons makes the
photon collider an ideal instrument for investigating the external 
spin-parity quantum numbers of particles \cite{r15D} such as Higgs bosons,
and for the study of CP violation effects \cite{r9}. 
%
\subsection{Luminosities and Cross sections}
The essential characteristics of the $e \gamma$ and $\gamma \gamma$
luminosity distributions can be determined by convoluting the photon
spectrum with the electron spectrum and with itself, respectively. \s

-- The {{$e \gamma$ luminosity}} distribution coincides with the
high-energy photon energy spectrum. Denoting the scaled $e \gamma$ invariant
mass by $m_{e\gamma} = M_{e\gamma}/\sqrt{s}$, it is given by
\begin{equation}
{d\mathcal{L}}/{dm^2_{e\gamma}} = F(y=m^2_{e\gamma})
\end{equation}
\\

\begin{figure}[h]
\begin{center}
\epsfig{figure=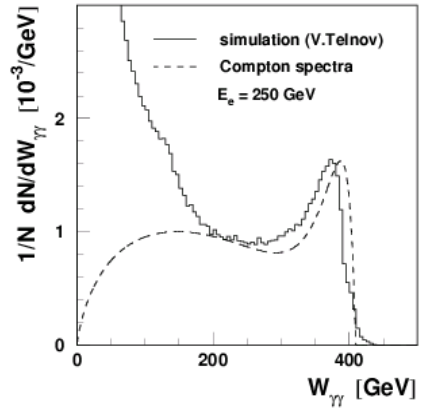,bbllx=3,bblly=3,bburx=250,bbury=233,width=5.85cm,clip=}
\caption{Luminosity distribution of the $\gamma \gamma$ collider; Ref.~\cite{r1}.}
\label{luminosity}
\end{center}
\end{figure}

-- The {{$\gamma \gamma$ luminosity}} distribution, on the other
hand, is determined by the standard self-convolution of the photon
spectrum. Denoting the $\gamma \gamma$ invariant mass by $m_{\gamma\gamma} 
= M_{\gamma \gamma}/\sqrt{s}$, the luminosity can be written as
\begin{equation}
{d\mathcal{L}}/{dm^2_{\gamma\gamma}} = \int_{m^2_{\gamma\gamma}}^{1}
               \frac{dy}{y} \,F(y) \,F(m^2_{\gamma\gamma} / y)
\end{equation}
If the helicities of the laser photons and the electrons are chosen
opposite to each other, a large fraction of the luminosity is accumulated
in the high-energy region, cf. Fig.\ref{luminosity}. For a polarization degree 
of 85\%
of the initial electrons and maximal polarization of the laser photons,
the values of the $e \gamma$ and $\gamma \gamma$ luminosities are
displayed in Tab.1 in units of $10^{34}$cm$^{-2}$s$^{-1}$.
\begin{table}[h]
\begin{center}
\begin{tabular}{|l|cc|}
\hline
\rule{0mm}{4mm}$\sqrt{s}_{ee}$ [GeV]  & 500  & 800        \\[0.1cm]
\hline
\hline
\rule{0mm}{4mm}$m_{e\gamma} \geq 0.8 m_{e\gamma}^{max}$           
& 0.9  & 1.3        \\[0.1cm]
\rule{0mm}{4mm}$m_{\gamma\gamma} \geq 0.8 m_{\gamma\gamma}^{max}$ 
& 1.1  & 1.7      \\[0.15cm]
\hline
\rule{0mm}{4mm}$e^+e^-$               & 3.4  & 5.8        \\
\hline
\end{tabular}
\caption{Luminosities for $e \gamma$ and $\gamma \gamma$ within a margin 
of size 0.2 below the maximum invariant energies, compared with the LC
$e^+e^-$ luminosity; Ref.~\cite{r1}.}
\end{center}
\end{table}
About one third of the corresponding $e^+e^-$ luminosity is accumulated
within a margin of 20\% below the maximum invariant energy:
\begin{equation}
{\mathcal{L}}_{\gamma\gamma}(\geq 0.8) \sim \frac{1}{3} {\mathcal{L}}_{ee}
\end{equation}
Since the cross sections for the $e \gamma$ and $\gamma \gamma$ processes
are significantly larger than the $e^+e^-$ annihilation cross sections,
the event rates are predicted of similar size for all three types of
collisions. \s

\begin{figure}[h]
\begin{center}
\epsfig{figure=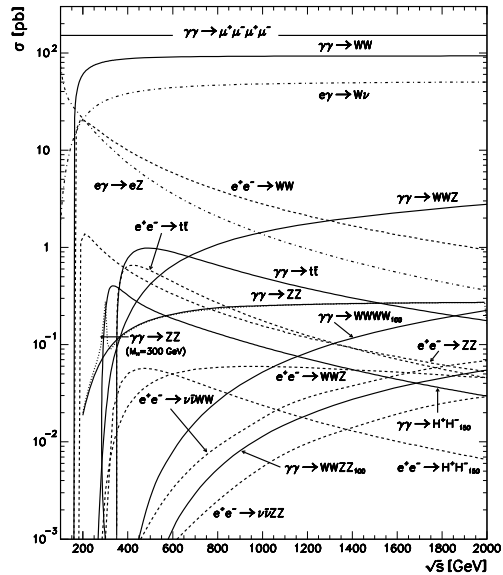,bbllx=3,bblly=3,bburx=531,bbury=656,width=6.6cm,clip=}
\caption{Cross sections for $e \gamma$ and $\gamma \gamma$ processes as functions of the energy; compared with $e^+e^-$ annihilation channels.}
\label{cxns}
\end{center}
\end{figure}

A sample of cross sections for $\gamma \gamma$ processes is collected in
Fig.\ref{cxns}. Since the size of the cross sections is large, extending from 
10 to $10^5$ fb, a large number of $10^3$ to $10^7$ events can be observed for 
a total integrated luminosity of 100 fb$^{-1}$ in the high-energy margin of 
the luminosity distribution.
%
\section{Electroweak Symmetry Breaking}
The discovery of the mechanism which breaks the electroweak symmetries and
the exploration of its nature are among the central experimental tasks at
the next-generation high-energy colliders. The Higgs mechanism is strongly
supported by high-precision analyses in the electroweak sector. However,
these measurements can still be interpreted within a large variety of models 
incorporating different realizations of the Higgs mechanism. They extend from 
the Standard Model to embeddings in supersymmetric theories up to theories of 
extra space dimensions. A photon collider can contribute to the task of uncovering 
the underlying structure. The Higgs coupling to two photons, measured by the 
size of the Higgs formation cross section in photon collisions, is sensitive 
to high scales in the theory. Moreover, photon collisions provide unique 
opportunities for the discovery of the heavy Higgs bosons in supersymmetric 
theories. Alternatively, strong electroweak symmetry breaking may be studied 
by measuring the static electromagnetic properties of the $W^\pm$ bosons in 
$\gamma \gamma$ fusion. These opportunities will be illustrated by a few 
typical examples. 
%
\subsection{Light Higgs in $\gamma \gamma$ Collisions}
The coupling of Higgs bosons to photons \cite{r16} is mediated by loops of
charged particles. In the Standard Model the main contributions are generated 
by top quark and $W$ boson loops. In scenarios beyond the Standard 
Model also heavier particles can contribute but the loops are suppressed with 
increasing masses according to the rules of quantum mechanics. [The 
suppression can be counter-balanced by rising Higgs couplings to the new 
particles if their masses are generated by the Higgs mechanism; however, for 
masses beyond $\sim 1/2$ TeV the theory would become strongly interacting and 
the perturbative loop argument would cease to be valid.]  \s

\begin{figure}[h]
\begin{center}
\epsfig{figure=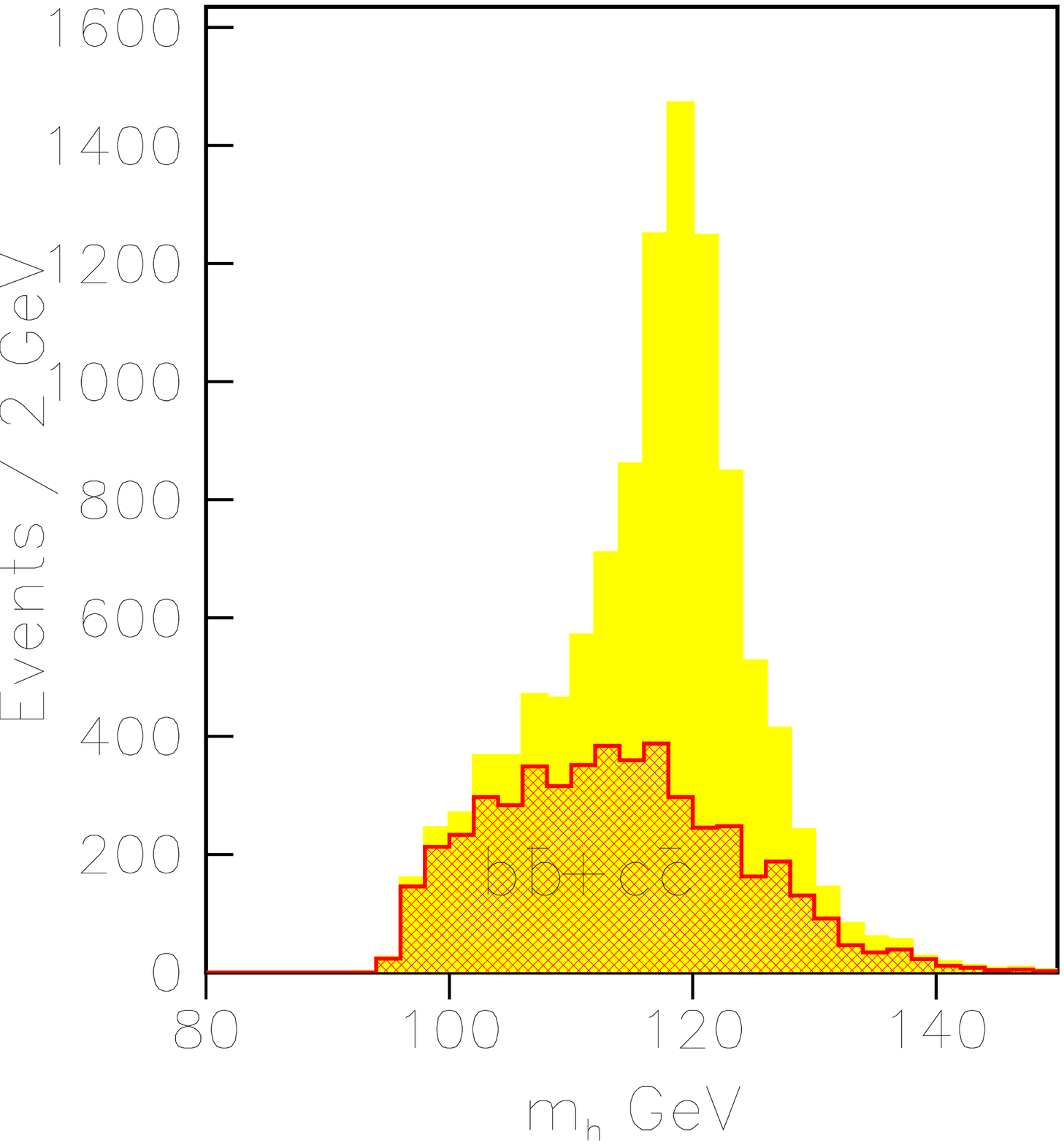,bbllx=1,bblly=12,bburx=556,bbury=612,width=5cm,clip=}
\epsfig{figure=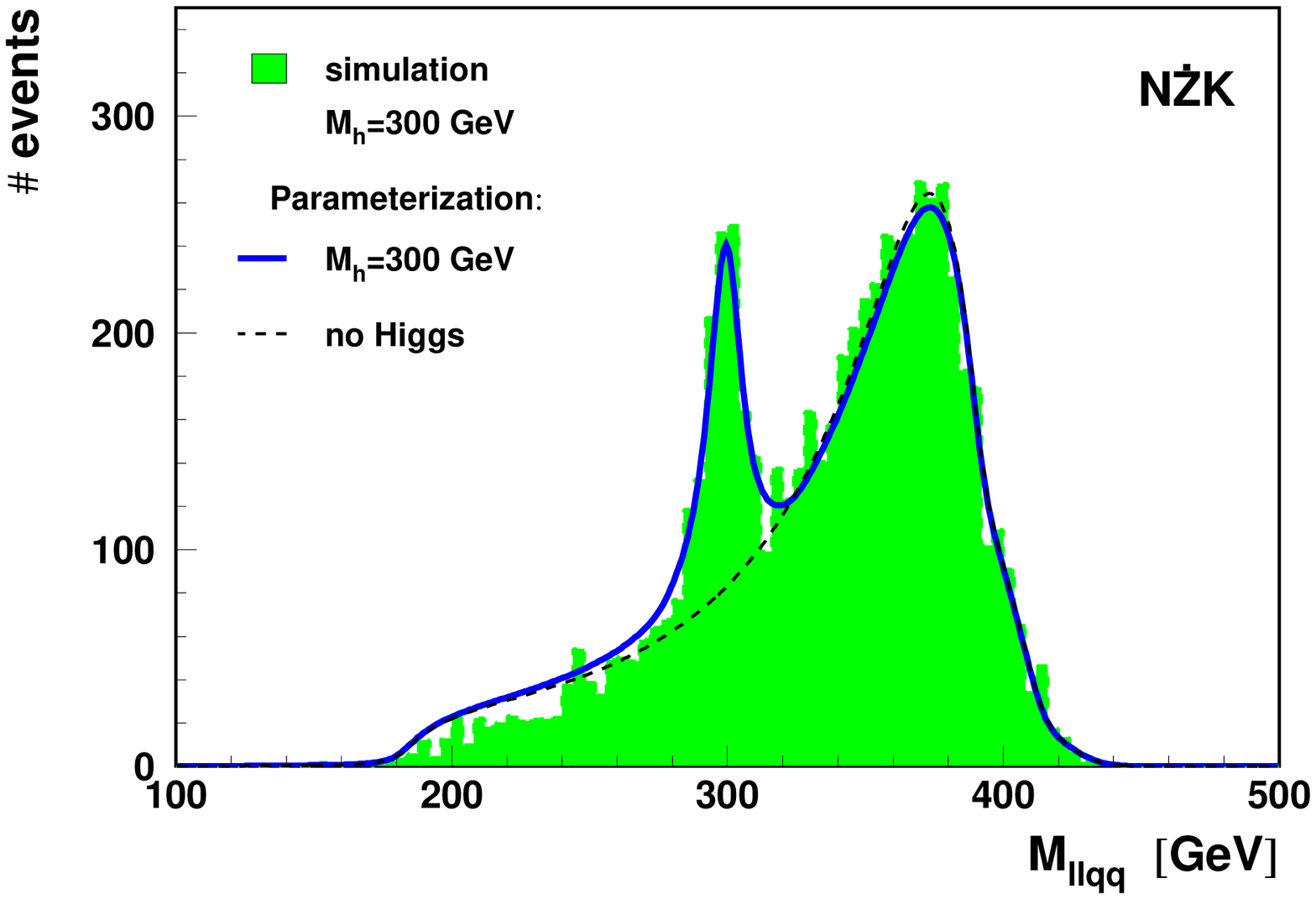,bbllx=1,bblly=-17,bburx=554,bbury=500,width=5cm,clip=}
\caption{Final state invariant $bb$, $ZZ$ masses for light and heavy Higgs boson
production in $\gamma \gamma$ collisions; Refs.~\cite{r4,r5}, respectively.}
\label{widthbbzz}
\end{center}
\end{figure}

The Higgs-$\gamma \gamma$ coupling determines the formation cross section
of Higgs bosons in $\gamma \gamma$ fusion. Apart from normalization, the
cross section can conveniently be expressed by the partial $\gamma \gamma$
decay width of the Higgs boson and the luminosity function; for narrow
Higgs bosons:
\begin{equation}
\sigma_{\gamma\gamma} \sim \Gamma_{\gamma\gamma} \,
                        {d\mathcal{L}}/{dm^2_{\gamma\gamma}} (M^2_H)
\end{equation}
Choosing the helicities of the photons opposite to each other enhances the
signal and helps suppress the background. The Higgs boson can be detected 
either as a peak in the invariant mass distribution, e.g. in $b\bar{b}$ or 
$ZZ$ decays, or by scanning, taking advantage of the sharp upper edge of the 
$\gamma \gamma$ spectrum. $b$ decays are leading below a Higgs mass of 
140 GeV; $Z$ decays, for which the background is strongly suppressed compared 
to $W$ decays, can be used for heavy Higgs masses, cf. Fig.\ref{widthbbzz}. 
In the lower mass region the partial width can be measured with an accuracy 
of 2.1\% \cite{r4}. \s

If the $H \to \gamma \gamma$ decay branching ratio is measured in 
Higgs-strahlung, the ratio $\tau_H = BR_{\gamma\gamma}/\Gamma_{\gamma\gamma}$ 
determines the lifetime $\tau_H$ of the Higgs boson. Depending on the error 
of the branching ratio \cite{rr20A}, an accuracy between 15 and 5\% may be reached, 
matching eventually the $WW$ channel. 
%
\subsection{Beyond the Standard Model (I)}
The high precision in the measurement of the Higgs-$\gamma \gamma$ coupling 
may be exploited to determine or, at least, constrain high scales in theories beyond 
the Standard Model in which the Higgs mechanism may be embedded. \s

-- In {2-Higgs doublet models} a situation could arise in which all the
properties of the lightest Higgs boson observed at LHC and ILC may be in 
concordance with SM expectations but the heavy Higgs bosons may not have been   
observed yet. In this situation the measurement of the $\gamma \gamma$ width 
can discriminate the extended Higgs model from the Standard Model \cite{r19}. 
Deviations of the $\gamma\gamma$ width from the Standard Model by more than a 
factor two could still be possible and they could easily be detected at a 
photon collider. \s

-- In models including {{triplet Higgs}} fields, doubly charged Higgs 
bosons will
be generated in addition to the singly charged Higgs bosons of 2-doublet
Higgs models. The double electric charge increases the production cross section 
by a factor 16 over the cross section for singly charged Higgs bosons. \s

\begin{figure}[h]
\begin{center}
\epsfig{figure=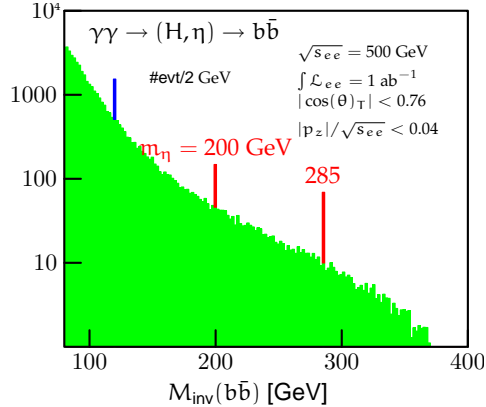,bbllx=154,bblly=400,bburx=568,bbury=752,width=6.3cm,clip=}
\caption{The formation of an axion-type particle $\eta$ of Little Higgs models in
addition to the Higgs particle in $\gamma \gamma$ collisions; Ref.~\cite{r21}.}
\label{littlehiggs}
\end{center}
\end{figure}

-- In {{Little Higgs models}} deviations from the Standard Model
are predicted across the interesting range of the scale parameter $f$
\cite{r20}. Moreover, the breaking of anomalous U(1) gauge symmetries may
generate new axion-type scalar particles \cite{r21} which can be searched
for, as narrow states, in channels parallel to the Higgs bosons, cf. 
Fig.\ref{littlehiggs}.  \s

-- Potentially large modifications of the Higgs-$\gamma \gamma$ coupling
are expected quite generally in models with {{strong electroweak 
symmetry breaking}}
\cite{r22}. If the photons couple directly to constituents of the Higgs
boson, the $\gamma \gamma$ branching ratio can rise to values significantly 
above the Standard Model value and may even become dominant, depending on the
size of the confinement parameters of the Higgs constituents. \s

-- In theories of {{extra space dimensions}}, the Kaluza-Klein states 
may alter the Higgs-$\gamma \gamma$ coupling \cite{r23}. In addition, in 
specific models, like Randall-Sundrum type (RS) models, the radion field may mix 
with the Higgs field and the properties of the Higgs particle may be modified
significantly; for details see e.g. Ref.~\cite{r24}. 
%
\subsection{Beyond the Standard Model (II): Supersymmetry Higgs Sector}
The contribution of charged supersymmetric particles to the loops in the 
Higgs-$\gamma \gamma$ coupling can give rise to noticeable effects, 
cf. Ref.~\cite{r25}, if the particles are not too heavy. \s

\begin{figure}[h]
\begin{center}
\epsfig{figure=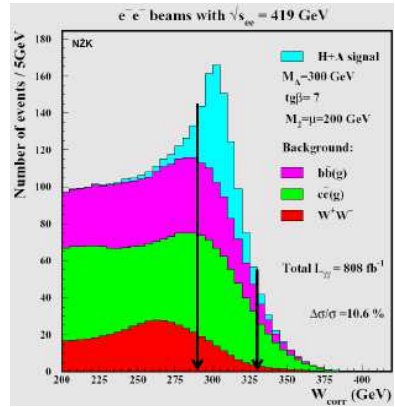,bbllx=2,bblly=3,bburx=237,bbury=250,width=5.1cm,clip=}
\caption{Experimental simulation of the production of heavy Higgs bosons
at a photon collider in supersymmetric theories \cite{r7}.}
\label{heavyhiggs}
\end{center}
\end{figure}

{{-- {{Heavy Higgs Bosons:}}}} If supersymmetry is realized in 
nature, a photon collider could be a unique instrument for exploring the 
Higgs sector of the theory. In the wedge centered around medium values of 
$\tan\beta$ beyond masses of about 200 GeV heavy Higgs bosons cannot be 
discovered at LHC and beyond 300 GeV even not at SLHC. Nor can the Higgs 
particles be discovered at the $e^+e^-$ linear collider beyond 250 GeV and 
500 GeV in the first phase and the second phase, respectively, as heavy 
scalar$+$pseudoscalar particles are produced in pairs. Thus only the lightest 
Higgs particle would be detected in standard channels while the other members 
of the Higgs spectrum would remain undiscovered. This wedge however can be 
covered in the $\gamma \gamma$ mode of the linear collider,
\begin{equation}
\gamma\gamma \to H,A
\end{equation}
which extends the mass reach in formation experiments to about 80\% of the 
total $e^+e^-$ energy \cite{r6}, i.e. up to 400 GeV and 800 GeV in the first 
and second phase of the operation. A clear Higgs signal can be isolated above 
the background \cite{r7} as demonstrated in Fig.\ref{heavyhiggs}. \s

Besides the discovery of heavy Higgs bosons, a photon collider could be
very valuable if not even unique in determining some of the parameters in
the Higgs sector of supersymmetric theories. \s

{{-- {{Higgs Mixing $\tan\beta$:}}}} 
The measurement of the
mixing parameter $\tan\beta$ is a difficult task for large values. Noticing 
that the Higgs Yukawa couplings to $\tau$ pairs are of the order of 
$\tan\beta$, the splitting of high-energy photons to $\tau$'s can be 
exploited to measure this parameter in $\tau\tau$ fusion, cf. Ref.~\cite{r26}:
\begin{equation}
\sigma[\gamma\gamma \to \tau\tau + h/H/A] \;\; \sim \;\; \tan^2\beta
\end{equation}
For moderate $A,H$ masses, the $h\tau\tau$ coupling is of the order of $\tan\beta$, 
while for large $A,H$ masses the size of the $A\tau\tau$ and $H\tau\tau$ couplings
is determined by this parameter; 
thus the entire mass parameter range can be covered. Since the splitting 
function of photons to leptons is hard, the energy of the $\tau$ beams is 
high so that Higgs particles with large masses can be produced. Introducing 
proper cuts to suppress the backgrounds, an accuracy of
\begin{equation}
\Delta\tan\beta = 0.9 \; {\rm{to}} \; 1.3
\end{equation}
can be expected across the entire medium to large $\tan\beta$ range \cite{r26}. \s

{{-- {{CP Violation:}}}} The 2-doublet Higgs sector of the
minimal supersymmetric extension of the Standard Model is automatically CP
conserving at the tree level. However, CP violation \cite{r9,r27} can be
induced by radiative corrections transmitting CP violating phases from the 
soft SUSY breaking Lagrangian to the Higgs system, in particular the relative 
phases between the higgsino mass parameter $\mu$ and the trilinear 
sfermion-Higgs parameter $A_f$. \s

CP asymmetries are naturally enhanced \cite{r9} in the decoupling regime
where $M_A \simeq M_H$. The near mass degeneracy of the scalar and
pseudoscalar states allows for frequent mutual transitions which induce large 
CP-odd mixing effects in CP violating theories. This is quantitatively 
described by the complex mixing parameter
\begin{equation}
\frac{1}{2} \tan 2\theta = \frac{\Delta^2_{HA}}
         {M^2_H - M^2_A - i [M_H\Gamma_H - M_A\Gamma_A]}
\end{equation}
where the off-diagonal CP violating parameter $\Delta_{HA}$ in the Higgs mass
matrix couples the two states. \s

The asymmetry between left- and right-handedly polarized top quark pairs,
produced in $\gamma\gamma$ fusion on the top and in the Breit-Wigner wings 
of the Higgs bosons, signals these CP violating effects \cite{r8}. \s

\begin{figure}[h]
\begin{center}
\epsfig{figure=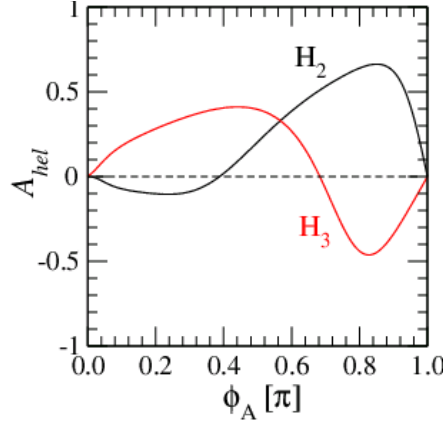,bbllx=5,bblly=3,bburx=237,bbury=227,width=5.8cm,clip=}
\caption{LR asymmetries predicted in CP non-invariant supersymmetric theories
on top of the two heavy Higgs bosons in $\gamma \gamma$ collisions; Ref.~\cite{r9}.}
\label{lrasymm}
\end{center}
\end{figure}

However, CP violation can be studied in a classical way by measuring 
asymmetries of inclusive cross sections between left- and right-polarized 
photons \cite{r9}:
\begin{equation}
A_{LR} = \frac{\sigma_{++} - \sigma_{--}} {\sigma_{++} + \sigma_{--}}
\end{equation}
Remarkably large asymmetries are predicted on top of the Higgs bosons in
$\gamma \gamma$ fusion, shown in Fig.\ref{lrasymm} as a function of the CP 
violating phase of the trilinear stop-Higgs coupling $A_t$. 
%
\subsection{Strongly Interacting $W$ Bosons}
If electroweak symmetry breaking is generated dynamically by new strong
interactions at a scale not far above 1~TeV, the longitudinal degrees of
the $W$ bosons, which are equivalent to the Goldstone bosons associated with 
spontaneous symmetry breaking, become strongly interacting particles.
As a result, the properties of the $W$ bosons will be affected by the
nearby new strong interactions. In particular deviations are expected in such a
scenario for the static electroweak parameters from the values which are generated
by the electroweak gauge interactions of the Standard Model:
\begin{eqnarray*}
 {\rm{magnetic}}\, W^\pm \,{\rm{dipole\, moment}} &:& \mu_W = 2 \times e/2M_W\\
 {\rm{electric}}\, W^\pm \,{\rm{quadrupole\, moment}} &:& Q_W = -e/M^2_W
\end{eqnarray*}
The deviations are described by linear combinations of parameters $\Delta
\kappa_\gamma$ and $\lambda_\gamma$ which modify the $WW\gamma$ vertex. \s

This modification can be studied \cite{r13} in the process
\begin{equation}
   \gamma \gamma \to W^+ W^-
\end{equation}
which receives large contributions from the t- and u-channel $W$ exchanges.
For a total energy of 800 GeV and an integrated luminosity of 1 ab$^{-1}$,
the expected sensitivity for the parameters $\Delta\kappa_\gamma$ and
$\lambda_\gamma$ is shown in Table 2, compared with the corresponding
sensitivities in the $e^+e^-$ mode.
\begin{table}
\begin{center}
\begin{tabular}{|r|cc|c|}
\hline
\rule{0mm}{4mm}& $\gamma\gamma$ : $J_z = 2$      &  $\gamma\gamma$ : $J_z = 0$  &
$e^+e^-$ : $J_z = 1$  \\[0.1cm]
\hline
\hline
\rule{0mm}{4mm}$\Delta \kappa_\gamma / 10^{-4}$  &   5.2  & 13.9 & 2.1   \\
\rule{0mm}{4mm}$\lambda_\gamma / 10^{-4}$        &   1.7  &  2.5 & 3.3   
\\[0.1cm]
\hline
\end{tabular}
\caption{Sensitivity of $WW$ pair production at $\gamma \gamma$ and $e^+e^-$ 
colliders to the $WW\gamma$ vertex parameters in theories of strong electroweak 
symmetry breaking; Ref.~\cite{r13}.} 
\end{center}
\end{table}
Apparently, the sensitivity on $\lambda_\gamma$ at the $\gamma \gamma$ collider 
is superior to the $e^+e^-$ collider, though not dramatically. 
%
\section{Supersymmetric Particles}
The genuine supersymmetric particle sector is the second domain 
of a photon collider for potentially
unique discoveries of particles which cannot be observed at 
other colliders. In supersymmetric scenarios in which sfermions are heavy but 
charginos/neutralinos light, cascade decays, the prime source of non-colored 
particles at LHC, do not include sleptons at a significant rate. On the other 
hand, the sleptons may be too heavy to be produced in pairs in the $e^+e^-$ 
mode of the linear collider. In this situation an $e \gamma$ collider could 
open the window to selectrons and $e$-sneutrinos \cite{r10,r11}. 
\begin{figure}[h]
\begin{center}
\epsfig{figure=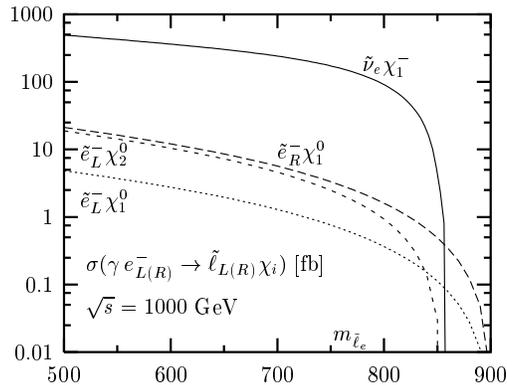,bbllx=158,bblly=526,bburx=387,bbury=703,width=6.7cm,clip=}
\caption{Total cross sections for the associated production of first generation
sleptons with charginos and neutralinos in $e^-\gamma$ collisions as functions
of the slepton masses; cf.~Ref.~\cite{r10}.}
\label{comp1000}
\end{center}
\end{figure}

While the energy in  the $e^+e^-$ mode must at least be larger than twice
the mass of the lightest selectron,
\begin{equation}
e^+ e^- \to \tilde{e}^+_i \tilde{e}^-_j  \;\;\; : \;\;\;
{\rm{min}}[m_{\tilde{e}}] \leq \frac{1}{2} \sqrt{s}
\end{equation}
the associated production of selectron and neutralino, or sneutrino and
chargino, in the $e \gamma$ mode,
\begin{equation}
e \gamma \to \tilde{e}_i \tilde{\chi}^0_j \;\;\ \mathrm{or} \;\;\ 
\tilde{\nu}_{e_i} \tilde{\chi}^\pm_j  
\;\;\; : \;\;\;
m_{\tilde{\ell}} \leq \sqrt{s} - m_{\tilde{\chi}}
\end{equation}
can give access to heavier selectrons and sneutrinos if the neutralinos
and charginos are light, Fig.\ref{comp1000}. \s

\begin{figure}[h]
\begin{center}
\epsfig{figure=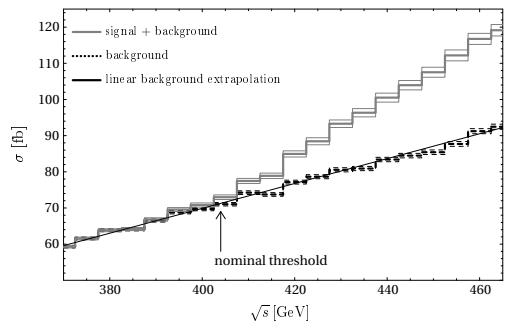,bbllx=25,bblly=380,bburx=490,bbury=681,width=6.6cm,clip=}
\caption{Sneutrino mass measurement by scanning the threshold in the process
$e \gamma \to \tilde{\nu} \tilde{\chi}$; Ref.~\cite{r11}.}
\label{snumass}
\end{center}
\end{figure}

Due to t-channel slepton exchanges the processes set in sharply 
at the thresholds proportional 
to the velocities $\beta$. The fast rise of the cross section 
can be exploited in threshold scans to 
determine the masses quite accurately. In the SPS1a$'$ example shown in 
Fig.\ref{snumass}, the sneutrino mass can be measured with an 
accuracy of 3~GeV, cf. Ref.~\cite{r11}. The 
accuracy is less than expected from chargino decays; however, all these 
experiments are quite involved due to numerous supersymmetry backgrounds 
to the supersymmetry signal so that cross checks are indispensable.
%
\section{Strong interactions and QCD}
%
\subsection{$\gamma \gamma$ Fusion to Hadrons}
One of the most demanding areas for the understanding of QCD as the
fundamental theory of the strong interactions is hadron production in
$\gamma \gamma$ collisions. Perturbative mechanisms are superimposed on
non-perturbative mechanisms with a difficult transition zone in between.
Since photons can fluctuate to vector mesons, the standard pomeron exchange
will contribute to the process. On the other hand, fluctuations to
quark-antiquark pairs at short distances will give rise to perturbative
QCD components, e.g. perturbative hard pomeron exchange and production of
mini-jets in $\gamma \gamma$ collisions. While for energies in the former 
LEP range, the relative weight of the mechanisms could not be determined
properly, the increase of the collision energy up to about 1~TeV 
at a photon collider is 
expected to discriminate between the various contributions. The theoretical 
picture is still evolving so that definite conclusions cannot be drawn yet at 
the present time \cite{r32}. 
%
\subsection{Quark-Gluon Structure of the Photon}
The measurement of the photon structure functions \cite{r33} at an $e
\gamma$ collider is of great theoretical interest. In contrast to the
proton structure function, the main characteristics of the photon structure
functions can be predicted theoretically. They are derived from the pointlike 
splitting of photons to quark-antiquark pairs \cite{r14} which gives rise to the
increase of the photon structure function $F_2$ for rising Bjorken-$x$ and 
to its uniform increase in the logarithm of the momentum transfer $\log Q^2$,
both features in sharp contrast to the proton. The QCD leading order
corrections \cite{r15A} modify the structure function in a characteristic
way. The perturbative radiation of gluons alters the $x$ dependence at 
$\mathcal{O}(1)$, though not overturning the characteristic rise in $x$, but it 
leaves the uniform $\log Q^2$ rise unchanged. This is a direct consequence
of asymptotic freedom \cite{r31A}; any other than the logarithmic $Q^2$ fall-off 
of the QCD coupling would have led to a power dependence of the structure
function. The non-logarithmic terms of the structure function receive
contributions from next-to-leading order of QCD \cite{r15B} but 
remnants from the vector-meson like component of the photon preclude 
the perturbative prediction of the absolute normalization for
non-asymptotic $Q^2$ values. Nevertheless, the theoretical prediction of the 
exceptional $x$ dependence and the uniform $\log Q^2$ rise render deep-inelastic
electron-photon scattering an exciting experimental task at an $e \gamma$ 
collider. \s

\begin{figure}[h]
\begin{center}
\epsfig{figure=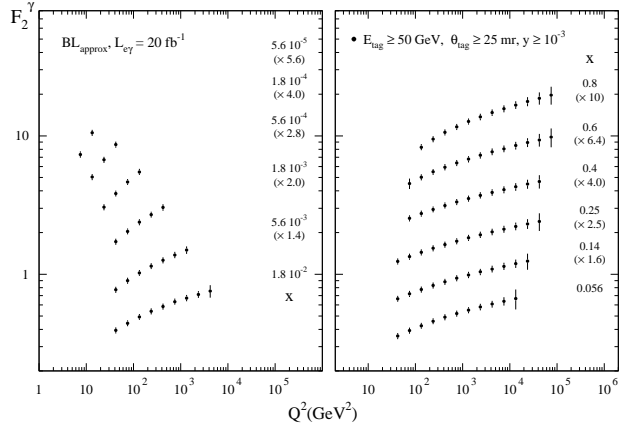,bbllx=58,bblly=82,bburx=530,bbury=761,width=5.7cm,clip=,angle=-90}
\caption{Accuracy with which the photon structure function $F_2$ can be
measured over a wide area in the $(x,Q^2)$ plane; Ref.~\cite{r35}.}
\label{f2measur}
\end{center}
\end{figure}

Neutral and charged-current mechanisms can be used to determine the
quark-parton content of the photon over a wide range of $Q^2$:
\begin{eqnarray}
e \gamma \to e X   \;\;&:&\;\; F_2^n \sim [4(u+c) +(d+s)] \nonumber\\
e \gamma \to \nu X \;\;&:&\;\; F_{2,3}^c \sim [(u+c) \pm (d+s)]
\end{eqnarray}
Range and accuracy with which $F_2$ can be measured are shown in 
Fig.\ref{f2measur}, Ref.~\cite{r35}. 
The analysis of charged-current deep-inelastic scattering allows the
separation of up- and down-quark components of the photon \cite{r34}.
\s

The vector-meson component of the photon as well as perturbative
gluon radiation suggest a gluon component within the photon. The
$(x,Q^2)$ dependence of the gluon 
component can be determined in two ways. First, the $\log Q^2$ evolution of the 
photon structure function is affected by the splitting of gluons to 
quark-antiquark pairs \cite{r35A}. Second, part of the high-transverse momentum 
jets in $e \gamma$ scattering,
\begin{equation}
e \gamma \to e + j + X
\end{equation}
are generated at the microscopic level by the subprocess
\begin{equation}
\gamma^\ast + g \to q + \bar{q}
\end{equation}
Studying the rapidity and transverse-momentum distributions of the jets
$j$ can thus be used to measure the gluon distribution of the photon
after the quark and anti-quark initiated Compton jets are subtracted \cite{r36}. 
\s

Finally, the spin structure function of the photon, $g_1(x,Q^2)$, can be
measured in deep-inelastic scattering of longitudinally polarized electrons
on circularly polarized photons. As shown earlier, circular 
polarization can be realized to a high degree for the photon beams. 
In next-to-leading order
the structure function $g_1$ can be predicted in the asymptotic regime,
the result matched closely by predictions based on standard evolution 
procedures \cite{r37}. 
%
\section{Summary}
A photon collider provides us with an experimental instrument with which
a large variety of problems across the entire range of physics beyond the Standard 
Model can be addressed. Moreover, within the Higgs sector and
the slepton sector of supersymmetric theories such a collider may give
unique access to heavy Higgs particles and heavy selectrons and
$e$-sneutrinos. Thus, by not only offering a platform for studying a rich 
experimental bouquet 
of interesting problems but also providing unique physics opportunities, a photon
collider can be considered a valuable component of a future linear
collider program.
%

\end{document}